\documentclass[english,aps,prl,twocolumn, superscriptaddress, amsmath]{revtex4}

\usepackage[T1]{fontenc}
\usepackage{physics}
\usepackage{textcomp}
\usepackage{color}
\usepackage{units}
\usepackage{amssymb}
\usepackage{graphicx}
\usepackage{esint}
\usepackage{bm}
\usepackage{natbib}
\usepackage{mciteplus}
\usepackage{ulem}
\usepackage{tabularx}
\usepackage{float}
\usepackage{gensymb}
\usepackage{xcolor}
\usepackage[colorlinks=true, citecolor=blue]{hyperref}
\usepackage{lipsum}
\usepackage{xfrac}

\usepackage{caption}
\usepackage{sidecap}  
\usepackage{babel}

\usepackage{multirow} 
\usepackage{dcolumn}

\newcommand{\vthres}{$V_{\rm TG}^{\rm th}$}

\begin{document}

\title {Spin-valley locking for in-gap quantum dots in a MoS{$\bm _2$} transistor}
  
\affiliation{Division of Physics and Applied Physics, School of Physical and Mathematical Sciences, Nanyang Technological University, Singapore 637371, Singapore}
\affiliation{School of Physics, University of New South Wales, Sydney, NSW 2052, Australia}
\affiliation{These authors contributed equally: Sangram Biswas, Radha Krishnan}

\author{Radha Krishnan $^{\P}$}
\author{Sangram Biswas $^{\P}$}
\affiliation{Division of Physics and Applied Physics, School of Physical and Mathematical Sciences, Nanyang Technological University, Singapore 637371, Singapore}
\affiliation{These authors contributed equally: Sangram Biswas, Radha Krishnan}

\author{Yu-Ling Hsueh}
\author{Hongyang Ma}
\author{Rajib Rahman}
\affiliation{School of Physics, University of New South Wales, Sydney, NSW 2052, Australia}

\author{Bent Weber}
\email{b.weber@ntu.edu.sg}
\affiliation{Division of Physics and Applied Physics, School of Physical and Mathematical Sciences, Nanyang Technological University, Singapore 637371, Singapore}

\begin{abstract}

 Spins confined to atomically-thin semiconductors are being actively explored as quantum information carriers. In transition metal dichalcogenides (TMDCs), the hexagonal crystal lattice gives rise to an additional valley degree of freedom with spin-valley locking and potentially enhanced spin life- and coherence times. However, realizing well-separated single-particle levels, and achieving transparent electrical contact to address them has remained challenging. Here, we report well-defined spin states in a few-layer MoS$ _2$ transistor, characterized with a spectral resolution of $\sim{50~\mu}$eV at ${T_\textrm{el} = 150}$~mK. Ground state magnetospectroscopy confirms a finite Berry-curvature induced coupling of spin and valley, reflected in a pronounced Zeeman anisotropy, with a large out-of-plane $g$-factor of ${g_\perp \simeq 8}$. A finite in-plane $g$-factor (${g_\parallel \simeq 0.55-0.8}$) allows us to quantify spin-valley locking and estimate the spin-orbit splitting ${2\Delta_{\rm SO} \sim 100~\mu}$eV. The demonstration of spin-valley locking is an important milestone towards realizing spin-valley quantum bits.
 
\end{abstract}


\maketitle

\section{Introduction}
Spins confined to semiconductor nanostructures have been investigated extensively for their potential to encode quantum bits (qubits) for quantum information processing \cite{hanson2008coherent, Burkard2023}. Atomically-thin semiconductors with hexagonal lattices \cite{xu2014spin, kormanyos2014spin, szechenyi2018impurity, pawlowski2021valley, pawlowski2019spin, altintacs2021spin}, such as bilayer graphene \cite{banszerus2021spin, eich2018spin} and the transition metal dichalcogenides (TMDCs) \cite{xu2014spin, kormanyos2014spin, szechenyi2018impurity, pawlowski2021valley, pawlowski2019spin, altintacs2021spin}, have recently attracted significant attention owing to their additional valley degree of freedom from non-equivalent conduction band minima at the $\rm K$ and $\rm {K^\prime}$ points (in graphene and TMDC monolayers) and $\rm Q$ and $\rm {Q^\prime}$ points (in TMDC multi-layers) of the Brillouin zone. TMDCs, in particular, offer strong spin-orbit coupling \cite{kormanyos2014spin} arising from the $d$-orbitals of the heavy transition metal atom, and a large tunable bandgap \cite{kormanyos2014spin}. In monolayers and odd-numbered multilayers, broken inversion symmetry combined with time-reversal symmetry can cause the spin and valley degrees to couple \cite{xu2014spin} at both K (monolayers) and Q (odd multi-layers) points, promising novel approaches to controlling spin-valley states \cite{kormanyos2014spin, szechenyi2018impurity, pawlowski2021valley, altintacs2021spin} with potentially enhanced spin life- and coherence times \cite{pawlowski2019spin,pawlowski2021valley}. 

\begin{figure*}[ht!]
\begin{center}
\includegraphics[width=1\textwidth]{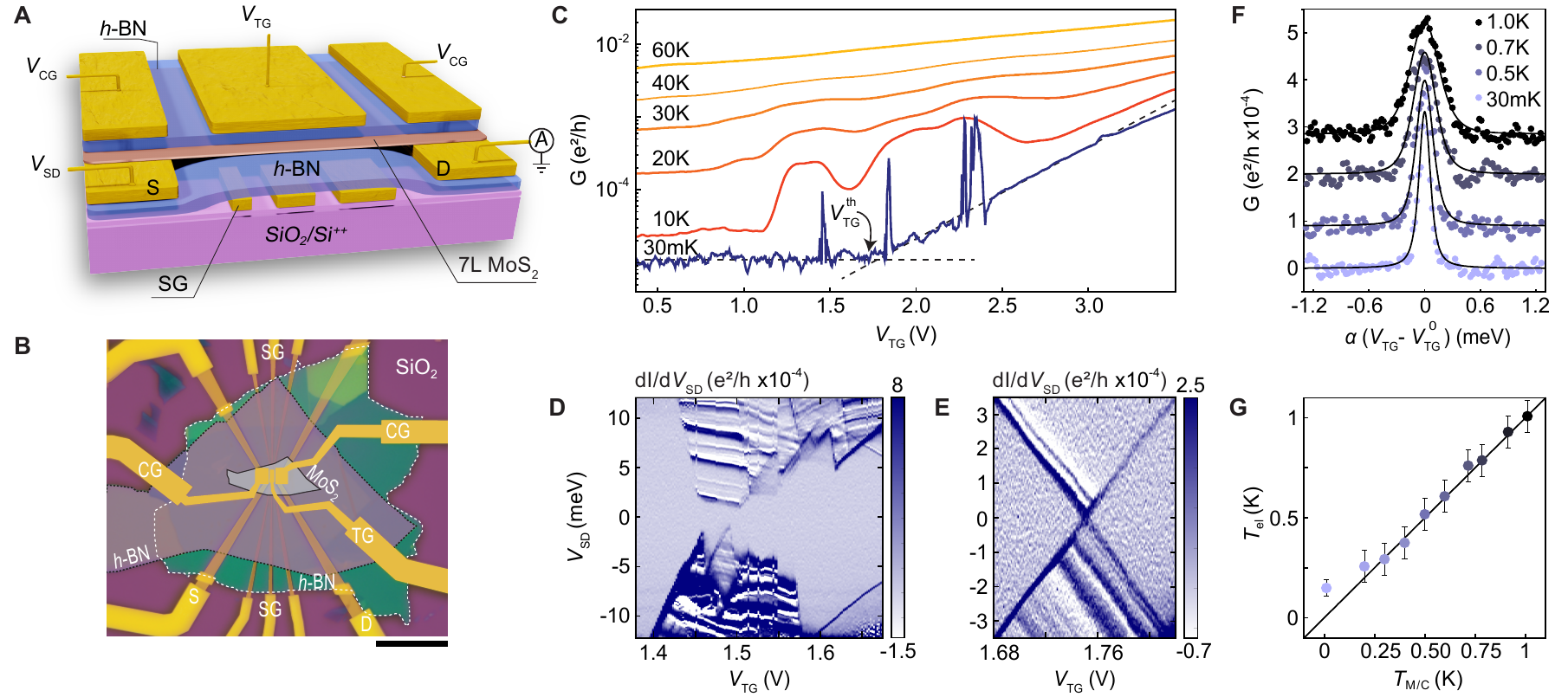}
\captionsetup{labelfont=bf,name={Figure},labelsep=period,justification=Justified}
\caption {{{Single-electron tunneling through in-gap states in a few-layer MoS$_{ 2}$ transistor. (A) A schematic of a $h$-BN encapsulated 7L MoS$_2$ transistor, contacted by source and drain Ohmic contacts. A top-gate (TG) is used to control the channel carrier density, while separate contact gates (CG) maintain a high carrier density in the contact regions even when the channel is near threshold. {(B)} An optical microscope image of the device, false-colored to indicate the individual layers of the device (Scale bar:~10~$\micro$m). {(C)} Channel transconductance curves measured at different temperatures. The intersection point of the dashed lines indicates the threshold voltage ({\vthres} = 1.78~V). {(D-E)} Finite-bias spectroscopy, recorded at a mixing chamber temperature of ${T_{\rm M/C}=30}$~mK, showing well-defined charge transitions. {(F)} Temperature-dependence of a Coulomb blockade peak (data in {(F)}), used to estimate the electron temperature (date offset for clarity). {(G)} Electronic temperature vs. mixing chamber temperature of our dilution refrigerator to estimate the lowest electron temperature achieved (${T_{\rm el} = (150 \pm 40)}$~mK)}.
\label{fig:dev}}}.
\end{center}
\end{figure*}


Consequently, there has been a significant effort to engineer TMDC-based quantum devices in which few or even single electronic charges and their spin can be isolated and controlled. Electrostatic confinement of electrons to quantum dots has so far been demonstrated in mono- and multi-layers of MoS$_2$ \cite{wang2018electrical,pisoni2018gate}, WS$_2$ \cite{song2015temperature}, and WSe$_2$ \cite{davari2020gate,boddison2021gate} and have shown clear signatures of Coulomb blockade in single-charge tunneling. Despite these significant advances in device engineering, however, it has remained challenging to isolate and address individual spin states in the few-electron regime \cite{hanson2007spins} to confirm predictions of their spin-valley character in electron transport spectroscopy. In part, this challenge arises from poorly screened disorder potentials at extremely low carrier density near threshold \cite{qiu2013hopping,jariwala2013band}, and a large electron effective mass \cite{pisoni2018gate}, dictating tight confinement potentials to within a few tens of nanometres \cite{davari2020gate}. Given these sizable challenges, only recently has transport through discrete energy levels been demonstrated in electrostatically defined TMDC quantum dots by Davari et al. \cite{davari2020gate}.

Different from electrostatically defined quantum dots, spins confined to atomic point defects or shallow dopants creating in-gap states \cite{koenraad2011single} have proven \cite{fuechsle2012single, weber2014spin} to provide tight confinement that is robust to disorder and yields well-separated single-particle levels with energy splitting of up to several meV \cite{fuechsle2012single,weber2014spin}. In-gap states have also been observed in TMDC-based quantum devices \cite{papadopoulos2020tunneling, devidas2021spectroscopy}. However, detailed investigations Zeeman anisotropy for different directions of the applied magnetic field, allowing to infer information on spin-valley character, have so far been missing. 

Resolving individual spin states and probing subtle changes in their state energy from applied electric or magnetic fields requires spectral resolution near the thermal limit ($3.5 k_{\rm B} T$). Yet, in-gap states are usually confined near or below threshold in semiconductor devices. In this extremely low carrier-density regime, it has remained challenging to form transparent electrical contacts \cite{wang2022making}, which stay Ohmic down to the required milliKelvin temperatures. 

Here, we overcome this problem by the use of gate-tunable contacts,  allowing us to perform transport spectroscopy with a resolution limited only by the electronic temperature of the reservoirs ($T_{\rm el}=150$~mK). Our detailed ground state magnetospectroscopy in vectorized magnetic fields confirms a pronounced Zeeman anisotropy with effective out-of-plane $g$-factor as large as $g_\perp \simeq 8$, consistent with spin-valley locking. We attribute a small but finite effective in-plane $g$-factor ($g_{\parallel} \simeq 0.55-0.80$) to inter-valley scattering in the presence of the defect potential, allowing us to estimate a spin-orbit splitting for electrons ($2\Delta_{\rm SO}\sim~$100$~\mu$eV). 

\section{Results and Discussion}
Figure~\ref{fig:dev}A-B shows the schematic of our MoS$_2$ transistor device alongside a false-colored optical microscope image. A seven atomic layers ($N = 7$) thick MoS$_2$ crystal (see supporting information) was encapsulated between two hexagonal boron nitride ($h$-BN) layers and contacted with Ti/Au source (S) and drain (D) electrodes. We use a separate metal top gate (TG) and contact gates (CG) to independently tune the electron density in the TMDC transistor channel near threshold, while maintaining a high carrier density within the contact regions. A set of local gate electrodes (SG), not used for the work presented, were left floating. 


Transfer curves measured at different temperatures are plotted in Figure~\ref{fig:dev}C, showing the typical $n$-type behaviour expected for natural MoS$_2$, in which doping arises from native point defects such as chalcogen (S) vacancies \cite{qiu2013hopping,hong2015exploring}. Near and below threshold (\vthres$\sim$1.78~V), we observe a series of sharp conductance resonances consistent with Coulomb blockade (CB) in single-charge tunneling. The finite conductance observed sub-threshold at higher temperatures can be shown (see supporting information) to arise from 2D variable range hopping \cite{qiu2013hopping,jariwala2013band} within impurity bands, consistent with shallow $n$-type doping. Bias spectroscopy (Figure~\ref{fig:dev}D) confirms signatures of stochastic Coulomb blockade \cite{Chuang2012}, with the detailed pattern of charge transitions in number and position varying from cool-down to cool-down. Conductance peaks near or above threshold have previously been attributed to point defect ensembles and/or disorder potentials affecting the conduction band tail  \cite{papadopoulos2020tunneling,devidas2021spectroscopy, ramezani2021superconducting, kotekar2019coulomb}. Sub-threshold peaks observed, however, may only be explained by tunneling through impurity-bound states within the bandgap \cite{papadopoulos2020tunneling,devidas2021spectroscopy}. 

As shown by the high-resolution close-ups in Figure~\ref{fig:dev}D-E, measured in an initial cool-down of the device, we observe single-electron tunneling through small clusters of in-gap states. However, we can also locate isolated charge transitions (Figure~\ref{fig:dev}E), which will be the focus of our investigation. For the transition shown, symmetric slopes bounding the Coulomb diamond indicate an approximately equal capacitive coupling to source and drain reservoirs. Multiple resonances with splitting of order meV, parallel to the diamond edges are seen, which can have a number of origins. At positive bias, we observe a single resonance above the ground state with splitting $\sim$300~$\mu$eV which may reflect a low-lying spin- or orbital excited state. Above, we observe a large gap in the excitation spectrum of at least $\sim$10~meV (see supplementary information Figure. S4). Assuming a lower limit of the orbital splitting $\Delta E \sim$10 meV we estimate the radius of the dot to be at most between 1.88 nm to 4.1 nm, depending on whether a 2D/3D dot or K/Q valleys are considered. These values are of similar order as what has been observed for S-vacancies in STM measurements \cite{vancso2016intrinsic}, confirming the atomic scale confinement of electrons in this device. Multiple parallel resonances at negative bias, with non-uniform spacing likely reflect a combination of excited states with features arising from a non-uniform reservoir density-of-states (DOS) \cite{papadopoulos2020tunneling}. The asymmetry in bias observed may suggest an asymmetry in the tunnel coupling to source and drain reservoirs \cite{escott2010resonant}.

An estimate of the effective temperature in the S and D electron reservoirs -- and hence our spectral resolution -- can be obtained by fitting the Coulomb peaks \cite{foxman1993effects} measured at a small bias of $V_{\text{SD}}=$ 0.01 mV, as shown in Figure~\ref{fig:dev}F (see supporting information for detail). Our fits confirm that tunneling proceeds through individual quantized energy levels \cite{beenakker1991theory,song2015temperature} rather than through a many-electron island in the classical regime \cite{kotekar2019coulomb}, as evident from the approximately linear increase in the peak resistance with temperature (see supplementary information). We extract $T_{\rm el} = (150 ~\pm~ 40)$~mK at the base temperature of our dilution refrigerator -- the lowest reported for Coulomb blockade signatures in TMDC  materials to date. The corresponding thermal broadening $3.5 k_{\rm B}T \sim~45~\mu$eV, together with a finite lifetime broadening $\Gamma\sim80~\mu$eV, sets the limit of our spectral resolution and attests to the quality of our contacts.

\begin{figure*}[ht!]
\centering
\includegraphics[width=0.95\textwidth]{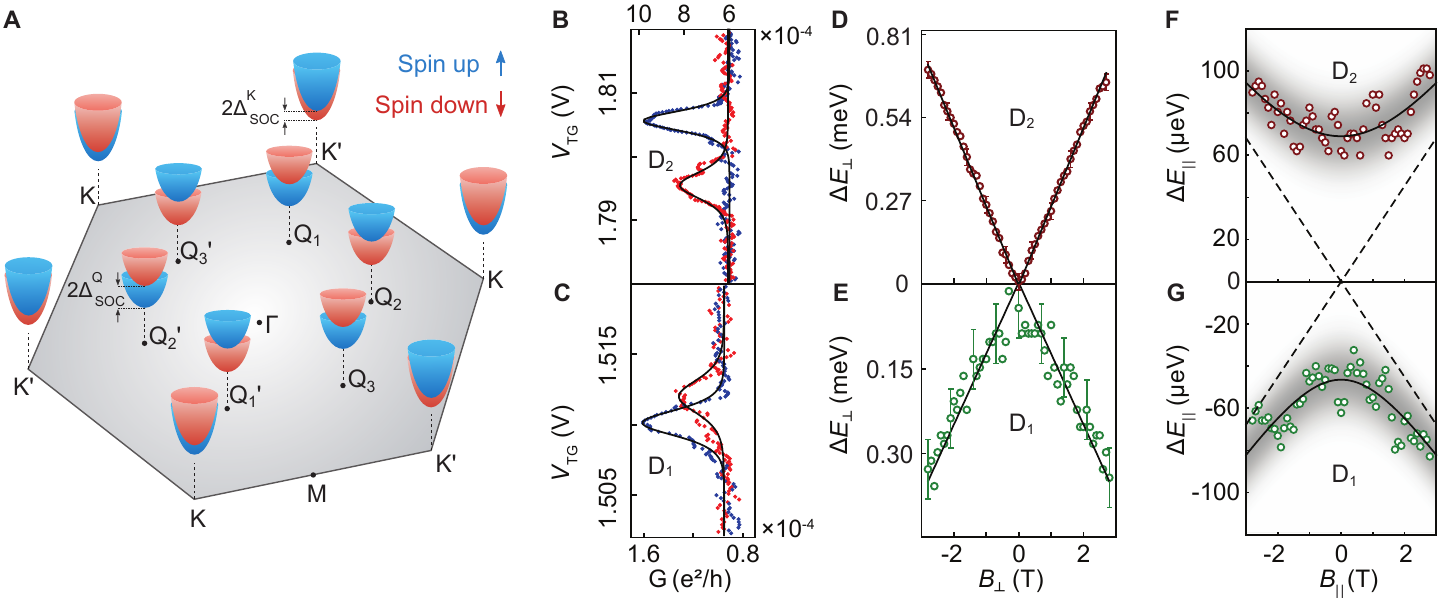}
\captionsetup{labelfont=bf,labelsep=period,justification=Justified}
\caption{{Spin-valley locking and Zeeman anisotropy for electron spins in few-layer MoS$_2$. (A) A schematic illustrating the spin–valley coupling for odd-layered MoS$_2$ at the conduction-band edges of the K- and Q-points of the hexagonal Brillouin zone. (B-C) Coulomb blockade peaks for transitions $D_1$ and $D_2$ at $B=0$ (red circles) and $B_{\perp}~=~2.8~$T (blue circles). Solid black lines are fits to the CB peak. (D-E) Zeeman shifted ground state transitions in $B_{\perp}$ for $\rm D_1$ (green circles) and $\rm D_2$ (maroon circles), respectively. (F-G) Shift of the ground state transitions in an in-plane magnetic field $B_{\parallel}$. Solid black lines are fits to Eq.~(\ref{Epara}), allowing us to extract the size of the spin-orbit splitting $2\Delta_{\rm SO}$. Shaded bands indicate the standard deviation (25$~\mu$eV) of the distribution of data points around their mean. Dashed lines are fits to Eq.~(\ref{Epara}) when $\Delta=0$.
\label{fig:g}}}.
\end{figure*}

\begin{figure*}[ht!]
\centering
\includegraphics[width=1\textwidth]{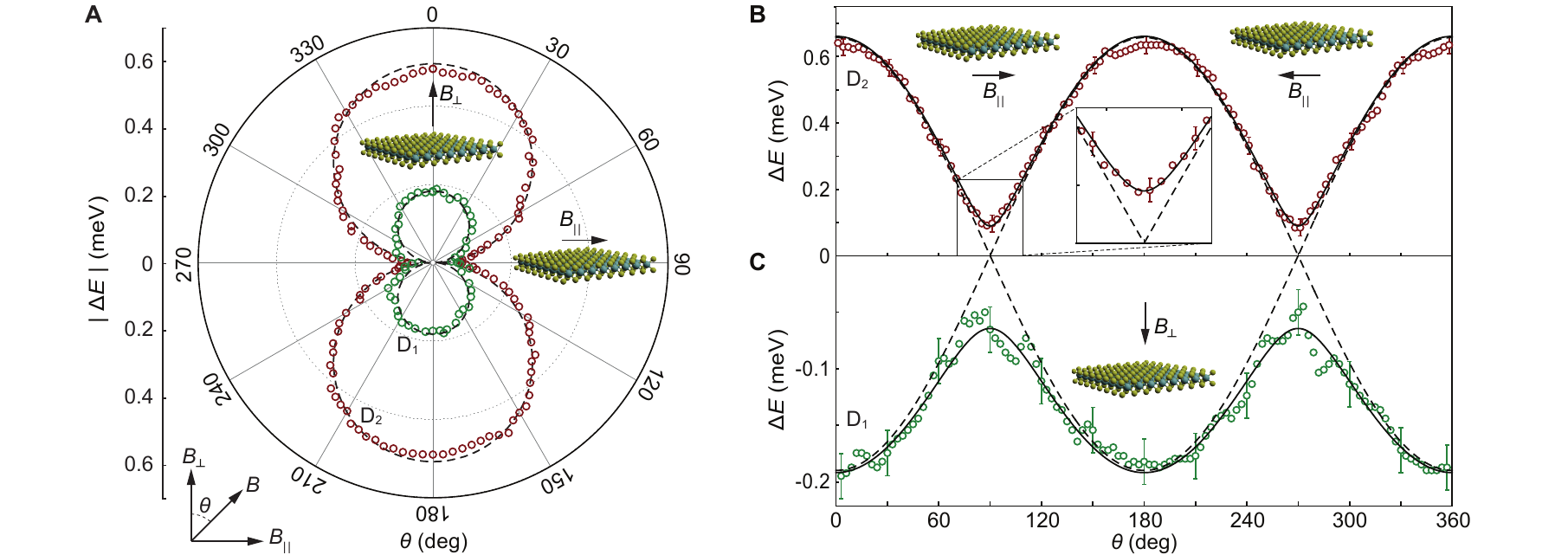}
\captionsetup{labelfont=bf,name={Figure},labelsep=period,justification=Justified}
\caption{The role of a finite in-plane $g$-factor. (A) Polar plot of the Zeeman anisotropy, extracted from the Coulomb peak shifts in a rotating magnetic field with angle $\theta$ between field vector and surface normal for both $\rm D_1$ (green open circles) and $\rm D_2$ (maroon open circles), measured at $|B|=2.8~$T. Dashed black lines are fits to equation Eq.~(\ref{Eperp}) assuming a purely out of plane Zeeman field, to determine $\left| \Delta E \right| = 0$ at $B_{\perp}=0$. Inserts indicate the magnetic field orientation for out-of-plane ($\theta =0\degree$) and in-plane ($\theta = 90\degree$) magnetic fields with respect to the MoS$_2$ atomic lattice (monolayer shown for clarity). (B-C) Zeeman shifts of the ground state transition for $D_1$ (green circles) and $D_2$ (maroon circles) as a function of magnetic field orientation ($\theta$). The insert shows a close-up near the in-plane direction of the magnetic field vector. The pronounced rounding confirms the presence of a finite in-plane $g$-factor. Solid black lines are fits to Eq.~(\ref{Eani}) without contribution from $\Delta$ to the anisotropy. Dashed lines are fits to Eq.~(\ref{Eperp}) to determine the $\Delta E = 0$ position.} \label{fig:gani}
\end{figure*}

Detailed insight into the interplay of spin and valley can be gained from ground state magnetospectroscopy and measurement of the Zeeman anisotropy, as presented in Figure~\ref{fig:g},\ref{fig:gani}. In MoS$_2$ multi-layers the conduction band (CB) if formed by degenerate minima at both the K- and the Q-points of the Brillouin Zone as shown in Figure~\ref{fig:g}A. For 7L crystals (see supplementary information), the CB minima is at Q-points followed by a mixing of K-points 137~meV above as confirmed from our DFT calculations. Shallow impurities such as sulphur vacancies can draw states from either conduction band pocket. In both TMDC monolayers and odd numbered multilayers broken inversion symmetry lifts the degeneracy of the valleys, split by 2$\Delta_{\rm SO}$ due to the intrinsic spin-orbit coupling \cite{kormanyos2014spin}. This gives rise to a non-zero Berry curvature of opposite sign in the respective Q/Q$^{\prime}$ and K/K$^{\prime}$ valleys, which acts like effective magnetic field in momentum space, giving rise to a valley magnetic moment, oriented perpendicular to the MoS$_2$ plane \cite{xu2014spin,schaibley2016valleytronics}. In addition, the spin-orbit coupling acts as an effective out-of-plane magnetic field, polarizing spins perpendicular to the MoS$_2$ crystal plane, ideally lacking an in-plane magnetic moment. Time-reversal symmetry dictates the spin splitting to have opposite sign at the K/K$^{\prime}$ and Q/Q$^{\prime}$ valleys, giving rise to an effective coupling between spin and valley pseudospin. Regardless of the valley occupation, given the spin texture in odd-numbered multilayers of MoS$_2$, spins are therefore expected to be locked to the out-of-plane direction with a vanishing Zeeman energy for in-plane magnetic fields. Observation of a pronounced Zeeman anisotropy is thus a clear sign of spin-valley locking. Indeed, spin-valley coupling of extended conduction band states at $\rm Q$ has been reported to dominate electron transport in odd-number multi-layers (trilayer and above) of MoS$_2$  \cite{wu2016even,roldan2014electronic}. In trilayer MoS$_2$, claims for both $\rm K$-valley \cite{pisoni2017gate} and $\rm Q$-valley 
 \cite{masseroni2021electron} transport have been made. However, the spin-valley nature of impurity bound states as evident from Zeeman anisotropy measurements has not yet been explored.

When a time-reversal symmetry breaking magnetic field is applied, the total Zeeman correction can be written as
\begin{eqnarray}
\Delta E_{\perp}&=&\dfrac{1}{2}g_\perp\mu_{\rm B} |B_{\perp}| + \Delta_{SO}
\label{Eperp}
\end{eqnarray}

\begin{eqnarray}
\Delta E_{\parallel} &=& \pm\dfrac{1}{2}\sqrt{ \big(g_{\parallel}\mu_{\rm B}B_{\parallel}\big)^2+(2\Delta)^2}
\label{Epara}
\end{eqnarray} 

where $g_{\perp} = s_z g_s+\tau g_{vl}$ is the effective out-of-plane $g$-factor, which can be expressed as a combination of spin ($g_s$) and valley ($g_{vl}$) $g$-factors, in which the valley ($\tau$) and spin ($s_z$) quantum numbers can take the values by $\tau = \{1, -1\}$ for \{$\rm Q/K$, $\rm Q^{\prime}/K^{\prime}$\} and $s_z = \{1, -1\}$ for \{$\uparrow$, $\downarrow$\}. $g_\parallel$ is the effective in-plane $g$-factor. $\Delta = \sqrt{\Delta_{\rm SO}^2 + \Delta_{\rm V}^2} $ is the inversion symmetry breaking term which can arise from a combination of spin-orbit coupling ($\Delta_{\rm SO}$) and inter-valley mixing $\Delta_{\rm V}$. $\Delta_{\rm SO}$ comprises of both $\tau_z$ and $s_z$ components, whereas $\Delta_{\rm V}$ consists of a $\tau_x$ component. Therefore, $\Delta_{\rm V}$ is added quadratically for $B_\parallel$ fields, while $\Delta_{\rm SO}$ is added quadratically in the case of $B_\parallel$ and linearly in the case of $B_\perp$ fields. This model is applicable to any unpaired valence spin states with spin-valley locking, without assumptions regarding charge occupation, shell filling sequence, or specific spin- and valley- g-factors, thus allowing the estimation of spin-orbit coupling strength through effective g-factor measurements. Figure~\ref{fig:g}B-E confirms Zeeman-shifted spin-valley ground states, measured for two charge transitions, $\rm D_1$ and $\rm D_2$, in an out-of-plane magnetic field, both measured in a second cool-down of the device (see supplementary information Fig.S4). For the out-of-plane Zeeman term (Eq.~\ref{Eperp}), the addition of a small $\Delta_{\rm SO}$ constitutes a constant offset to the Zeeman term, i.e. with no change in slope and hence $g_{\perp}$. From fits to a linear Zeeman field, we extract out-of-plane effective $g$-factors $g_\perp$=$2.2\pm0.2$ ($\rm D_1$) and $g_\perp$= $8.1\pm0.2$ ($\rm D_2$), respectively, where we have assumed gate levers of $\alpha_1 = 76\pm4~$meV/V and $\alpha_2 = 62\pm4~$meV/V, extracted from the slope of the CB diamonds (see supporting Information).

Ideally, the out-of-plane polarized spins in odd-layered TMDCs are expected to couple weakly to any in-plane magnetic fields. However a small but finite Zeeman shift in $B_{\parallel}$-fields can be detected as shown in Figure~\ref{fig:g}F,G. The dashed black lines indicate the corresponding Zeeman slope for $\Delta=$0.
Taking into account an inversion symmetry-breaking term $\Delta$ in Eq.~\ref{Epara}, we can fit the data (solid black line in Fig.~\ref{fig:g}F-G). The term $\Delta$ gives rise to departure from a linear in-plane Zeeman shift at low $B_{\parallel}$. At $B_{\parallel}\sim$2~T the in-plane field is strong enough to cant the spins into the MoS$_2$ crystal plane giving rise to a linear Zeeman shift for fields above. From fits to Eq.~(\ref{Epara}), we extract $g_{\parallel}=0.8\pm0.1$ for both transitions. It also allows us to extract $2\Delta=138\pm6~\mu$eV ($\rm D_2$) and $2\Delta=93\pm6~\mu$eV ($\rm D_1$), just within the spectral resolution limit of our measurement ($80~\mu$eV at 150~mK), and of similar magnitude as the low lying excited state at positive bias observed in Figure.~\ref{fig:dev}E. Although, there could be other sources of symmetry-breaking, such as Rashba spin-orbit coupling from field-effect gating, it is expected to be non-negligible only in aggressively gated devices, for example, dual-gated or devices with ionic liquid gating \cite{piatti2018multi}. Since here, the transitions measured are sub-threshold, at low top-gate voltages, we expect such effects to play a minor role in our device.

We confirm the notion of a pronounced Zeeman anisotropy in Figure~\ref{fig:gani}A, measured at $|B| = 2.8$~T for both D$_1$ and D$_2$, demonstrating near-vanishing Zeeman shifts for in-plane magnetic fields at angles $\theta = 90$\degree~and 270\degree~from the surface normal. Reasonable fits can indeed be obtained by assuming $B_{\perp}= |B| \mathrm{cos}\theta$ (dashed black lines in Figure~\ref{fig:gani}A), assuming that magnetic moments are strictly spin-valley locked out-of-plane. These fits also allows us to determine the reference point where $\left| \Delta E \right| = 0$, that is, where the out-of-plane Zeeman field vanishes in the absence of an in-plane $g$-factor (compare dashed lines in Figure~\ref{fig:gani}B-C). 

Out-of-plane Zeeman shifts have been demonstrated \cite{papadopoulos2020tunneling, davari2020gate}, in transport through both defects \cite{papadopoulos2020tunneling} and electrostatically-defined quantum dots \cite{ davari2020gate}, reporting out-of-plane $g$-factors ranging from 3.4-15.8 \cite{papadopoulos2020tunneling} and 0.8-2.4 \cite{ davari2020gate}, respectively. While we cannot fully exclude the possibility that the out-of-plane $g$-factor of $\rm D_1$ could represent a pure spin state without any valley-Zeeman contribution \cite{tsunetomo2021spin}, the large $g$-factor observed for $\rm D_2$, and the pronounced $g$-factor anisotropy observed in both transitions can only be explained by a combination of spin and valley Zeeman effects. Further extracting the respective contributions of spin- and valley $g$-factors to the total effective $g$-factor would require knowledge of the precise charge occupation and spin-valley eigenspectrum, neither of which are known with certainty. Although an orbital magnetic moment could in principle contribute to the overall enhancement of the $g$-factor \cite{davari2020gate}, we expect them to play a minor role in the atomically confined spin states \cite{pryor2006lande} considered here.

Consistent $g_{\parallel}$ can also be confirmed directly from Zeeman shifts of the ground state transition for D1 and D2 as a function of magnetic field orientation ($\theta$), as summarized in Figure~\ref{fig:gani}B-C. The dashed black lines in Figure~\ref{fig:gani}B-C are the same out-of-plane Zeeman fits as determined earlier in Figure~\ref{fig:gani}A, scaled by the lever arms, assuming $g_\parallel =0$. In this representation, the fits are seen to clearly underestimate the measured data at in-plane angles $\theta = 90\degree$ and $\theta = 270\degree$ for which a finite in-plane Zeeman shift is present. A better fit to the data, including the pronounced rounding at the minima, can be obtained by considering a finite in-plane $g$-factor such that

\begin{eqnarray}
\Delta E &=& \pm\dfrac{1}{2}\sqrt{ (g_{\perp }\mu_{\rm B}B\cos{\theta})^2+(g_{\parallel}\mu_{\rm B}B\sin{\theta})^2}\notag \\
\label{Eani}
\end{eqnarray} 

As shown by the solid black lines, fits to Eq.~(\ref{Eani}) allow us to confirm comparable values, $g_{\perp}=2.2\pm0.1$ ($\rm D_1$), $g_{\perp}=7.7\pm0.2$ ($\rm D_2$), $g_{\parallel}=0.6\pm0.1$ ($\rm D_1$), and $g_{\parallel}=0.7\pm0.2$ ($\rm D_2$) for the two transitions, in agreement with those extracted from the in-plane Zeeman shifts in Figure~\ref{fig:g}F-G. We note that contributions from $\Delta$ in Eq.~(\ref{Eani}) are not detectable within the fit quality, since the measurement was taken at a large enough magnetic field magnitude $|B|=$2.8 T, at which the Zeeman shifts are nearly linear in both in-plane and out-of-plane magnetic fields. The rounding at the minima therefore arises purely from the presence of a finite in-plane $g$-factor.

A finite in-plane $g$-factor as observed can indeed be expected to arise from valley-mixing in the presence of an abrupt defect potential \cite{szechenyi2018impurity},
\begin{equation}
    g_{\parallel}=g_s\dfrac{\Delta_{\rm V}}{\Delta_{\rm SO}}
\end{equation}

As seen from the expression above, $g_\parallel$ is expressed through the ratio of inter-valley scattering and SOC splitting and can hence be used to quantify the degree of spin-valley locking. For $\Delta_{\rm V} \simeq \Delta_{\rm SO}$, an isotropic spin $g$-factor would be expected. However, if scattering between valleys is negligible ($\Delta_{\rm V} =0$), electron spin is locked to a well-defined valley degree of freedom and $g_\parallel =0$ is recovered. Here, we encounter an intermediate scenario where a pronounced $g$-factor anisotropy represents spin-valley locking, but a finite in-plane $g$-factor arises from inter-valley scattering.   

In monolayers of MoS$_2$, upper limits to $\Delta_{\rm V}$ of 150~$\mu$eV and 1.5~meV have been predicted \cite{szechenyi2018impurity} for S and Mo vacancies, respectively. Considering also the predicted spin-orbit splitting $\Delta_{\rm SO} = 1.5$~meV for the conduction band in monolayers \cite{kormanyos2014spin}, as well as $g_s \sim 2$ predicted by theory \cite{kormanyos2014spin}, we expect $g_{\parallel} \approx 0.3$ for S vacancies, while a nearly isotropic $g_s \sim g_\parallel$ would be expected for Mo vacancies. Although both estimates are likely to be modified in multi-layers, their ratio agrees reasonably well with that expected for S vacancies. While we cannot fully eliminate the role of disorder potentials in confining individual spins, S-vacancies are likely candidates given that they are prevalent in natural MoS$_2$ crystals \cite{tumino2020nature,hong2015exploring}, and are known as shallow $n$-type dopants \cite{qiu2013hopping, schuler2019large}. Strong $n$-type behaviour is indeed reflected in the transfer curves of our device, with in-gap states observed close to the conduction band edge (Figure~\ref{fig:dev}C), similar to the case of shallow donors in covalent semiconductors such as silicon or GaAs. Thus, we expect the defect wave function to inherit the characteristics of the conduction band electronic structure \cite{Kohn1955}. The relatively large ratio $\Delta_{\rm SO}/\Delta_{\rm V}~\approx 3.5$ of spin-orbit coupling to inter-valley scattering inferred from our experiments further indicates that the spin-orbit energy scale dominates the sum of their squares $\Delta^2 = \Delta_{\rm SO}^2 + \Delta_{\rm V}^2$, allowing us to estimate $2\Delta_{\rm SO} \approx 2\Delta \approx 100~\mu$eV. Although low compared to calculations \cite{kormanyos2014spin} and measurements in monolayers \cite{marinov2017resolving}, this value possibly reflects the reduced SOC strength expected in strongly confined in-gap states and in multi-layers \cite{chang2014thickness}. Further suppression of the SOC gap has also been predicted for quantum dots with reduced radius \cite{bieniek2020effect}, similar to what would be expected in  highly-confined in-gap wave functions, but different from the soft confinement potentials encountered in electrostatically defined quantum dots of larger radii. It is hard to predict what would be expected for defect-bound spin-valley states in MoS$_2$. 

We note that in bilayer graphene \cite{eich2018spin, banszerus2021spin}, similar spin-valley locked ground states have recently allowed the demonstration of valley Zeeman splittings with $g$-factors as large as $g \sim 120$ \cite{lee2020tunable}, and with a spin-orbit gap of $\approx 60~\mu$eV \cite{banszerus2021spin}. The larger spin-valley splitting due to stronger SOC achievable in MoS$_2$ monolayers, however, gives rise to the strong suppression of in-plane magnetic moments in mono- and odd-layered MoS$_2$, different from the case of graphene in which $g_\parallel \approx 2$ has been reported \cite{eich2018spin}. An aspect to consider for future work is to investigate the role of gate tunable spin-orbit coupling, similar to what has recently been demonstrated for quantum point contacts in bilayer graphene \cite{banszerus2020observation}.

\section{Conclusion}
To conclude, we have demonstrated single-electron tunneling through in-gap states near threshold of a few-layer MoS$_2$ transistor. From sensitive ground state magnetospectroscopy near the thermal resolution limit, we confirm spin-valley locking of well-defined spin states as detected from a pronounced anisotropy of the Zeeman splitting in vectorized magnetic fields. The finite in-plane $g$-factor observed has allowed us to extract an estimate for the spin-orbit splitting in the conduction band of order $2\Delta_{\rm SO} \approx 100~\mu$eV. 
The observation of spin-valley locking resolved via the Zeeman anisotropy of well-defined spin states in a MoS$_2$ transistor device for the first time is an important milestone towards determining spin life- \cite{weber2018spin} and coherence \cite{pla2012single} times in TMDC-based spin-valley quantum bits.  

\section{Acknowledgements}

This research is supported by the National Research Foundation (NRF) Singapore, under the Competitive Research Programme ``Towards On-Chip Topological Quantum Devices'' (NRF-CRP21-2018-0001), with further support from the Singapore Ministry of Education (MOE) Academic Research Fund Tier 3 grant (MOE2018-T3-1-002) ``Geometrical Quantum Materials''. BW acknowledges a Singapore National Research Foundation (NRF) Fellowship (NRF-NRFF2017-11). The device fabrication in the work was carried out at the Micro and Nano-Fabrication Facility (MNFF), Centre of Advanced 2D Materials (CA2DM) at the National University of Singapore.

\section{Author Contributions}
${\P}$~RK and SB contributed equally to this work. RK carried out the device nanofabrication. SB performed the electrical measurements with the help of RK. RK, SB, and BW analyzed the data with input by YH and RR on crystal symmetry and $g$-factor anisotropy. HM performed the DFT calculations. BW perceived and coordinated the project. RK, SB, and BW wrote the manuscript with input from all authors. 

\bibliography{References}

\end{document}